# The CMS Particle Flow Algorithm

**Florian Beaudette on behalf of the CMS collaboration***
*Laboratoire Leprince-Ringuet, IN2P3-CNRS*
*E-mail:* Florian.Beaudette@cern.ch

A particle flow event-reconstruction algorithm has been successfully deployed in the CMS experiment and is nowadays used by most of the analyses. It aims at identifying and reconstructing individually each particle arising from the LHC proton-proton collision, by combining the information from all the subdetectors. The resulting particle-flow event reconstruction leads to an improved performance for the reconstruction of jets and MET, and for the identification of electrons, muons, and taus. The algorithm and its performance will be described. The commissioning phase, during which it was demonstrated that the algorithm was performing as expected from the simulation up to a high level of precision, will be presented. Finally, a selection of recent improvements in the CMS analyses obtained thanks to the particle-flow algorithm will be discussed.

*Speaker.

## 1. The Particle Flow Algorithm

The Particle Flow (PF) algorithm aims at identifying and reconstructing all the particles from the collision by combining optimally the information of the different subdetectors. The ingredients for an efficient PF algorithm arise from these principles: to maximise the separation between charged and neutral hadrons, a large field integral and a good calorimeter granularity are of primary importance. An efficient tracking is a key item as well as a material budget in front of the calorimeters as small as possible. The CMS detector [1] fulfill several of these conditions with a field integral more than twice larger than in other past or existing experiments; an electromagnetic calorimeter with an excellent resolution and granularity, and a tracker system fully exploited by the iterative tracking algorithm.

The CMS PF algorithm [2] relies on a efficient and pure track reconstruction, on a clustering algorithm able to disentangle overlapping showers, and on an efficient link procedure to connect together the deposits of each particle in the sub-detectors. Simplifying, the algorithm can be described as follows; a detailed description can be found in [2]. The tracks are extrapolated through the calorimeters, if they fall within the boundaries of one or several clusters, the clusters are associated to the track. The set of track and cluster(s) constitute a charged hadron and the building bricks are not considered anymore in the rest of the algorithm. The muons are identified beforehand so that their track does not give rise to a charged hadron. The electrons are more difficult to deal with. Indeed, due to the frequent Bremsstrahlung photon emission, a specific track reconstruction [3] is needed as well as a dedicated treatment to properly attach the photon clusters to the electron and avoid energy double counting. Once all the tracks are treated, the remaining clusters result in photons in case of the electromagnetic calorimeter (ECAL) and neutral hadrons in the hadron calorimeter (HCAL).
Once all the deposits of a particle are associated, its nature can be assessed, and the information of the sub-detectors combined to determine optimally its four-momentum. In case the calibrated calorimeter energy of the clusters, which is simply a linear combination of the ECAL and HCAL energy deposits, associated to a track is found to be in excess with respect to the track momentum at more than one sigma, the excess is attributed to an overlapping neutral particle (photon or hadron), carrying an energy corresponding to the difference of the two measurements.
The resulting list of particles, namely charged hadrons, photons, neutral hadrons, electrons and muons, is then used to reconstruct the jets, the missing transverse energy ($E_{\mathrm{T}}^{\mathrm{miss}}$), to reconstruct and identify the taus from their decays products and to measure the isolation of the particles.

The performance of the algorithm have been first been studied with simulated events [2]. About 90% of the jet energy is carried by the charged hadrons and the photons the energies of which are measured with a high precision by the tracker and the ECAL respectively, while the remaining 10%, carried by the neutral hadrons, is measured with the hadron calorimeter (HCAL) with a $120\%/\sqrt{E}$ resolution. As visible in Fig. 1, where the jets reconstructed with the sole calorimeters and the jets reconstructed with the PF candidates are compared for di-jets events in the barrel, between 95% and 97% of the jet energy, depending on the jet $p_{\mathrm{T}}$, is reconstructed in the case of the PF compared to 60%-80% for the calorimeter jets (left plot). In addition, the gain in resolution can be up to a factor of 3 (right plot). Furthermore, the angular resolution is improved by a factor 2-3.

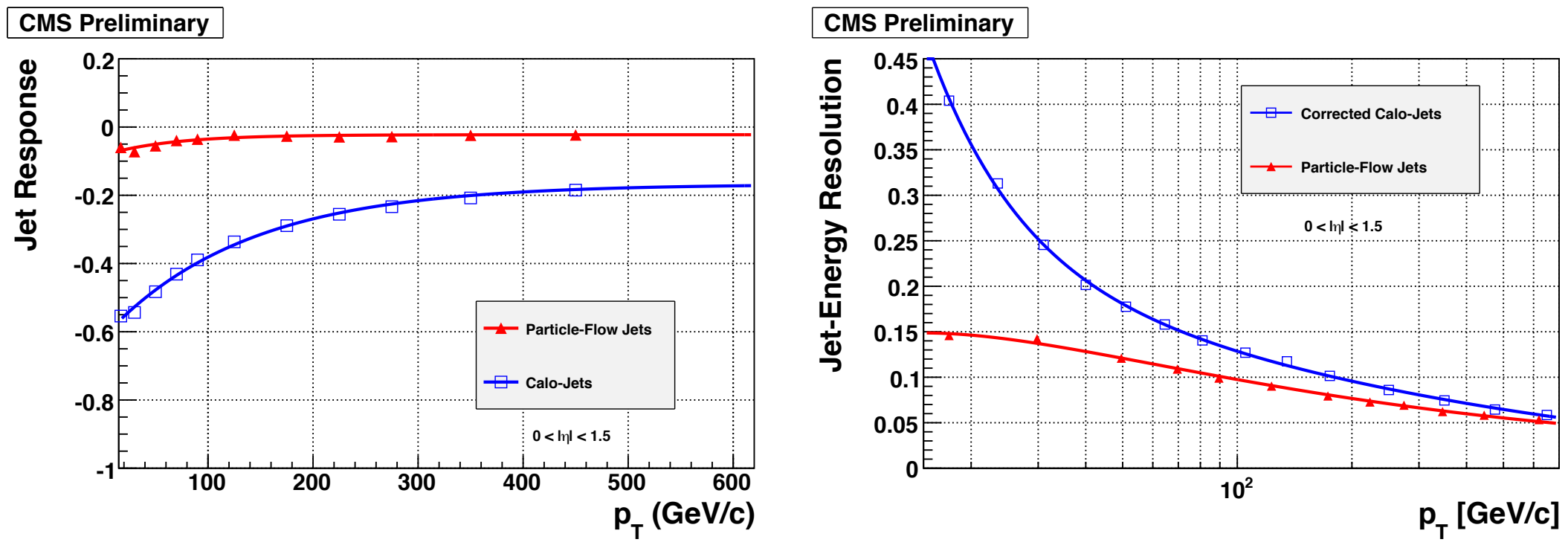

**Figure 1:** (*left*) Jet response for di-jets events in the barrel reconstructed with the particle flow (red triangles) and the calorimeters (blue open squares). (*right*) Jet resolution for di-jets events in the barrel reconstructed with the particle flow (red triangles) and the calorimeters (blue open squares).

The hadronic decays of the $\tau$-leptons result in narrow jets made of a small number of particles, and do not contain neutral hadrons. Consequently, the reconstruction of the $\tau$-leptons is highly improved by the PF reconstruction. Firstly, the measurement of the $\tau$ visible energy as shown in Fig. 2 is spectacularly enhanced; secondly the identification of the particles provides access to the decay channel.

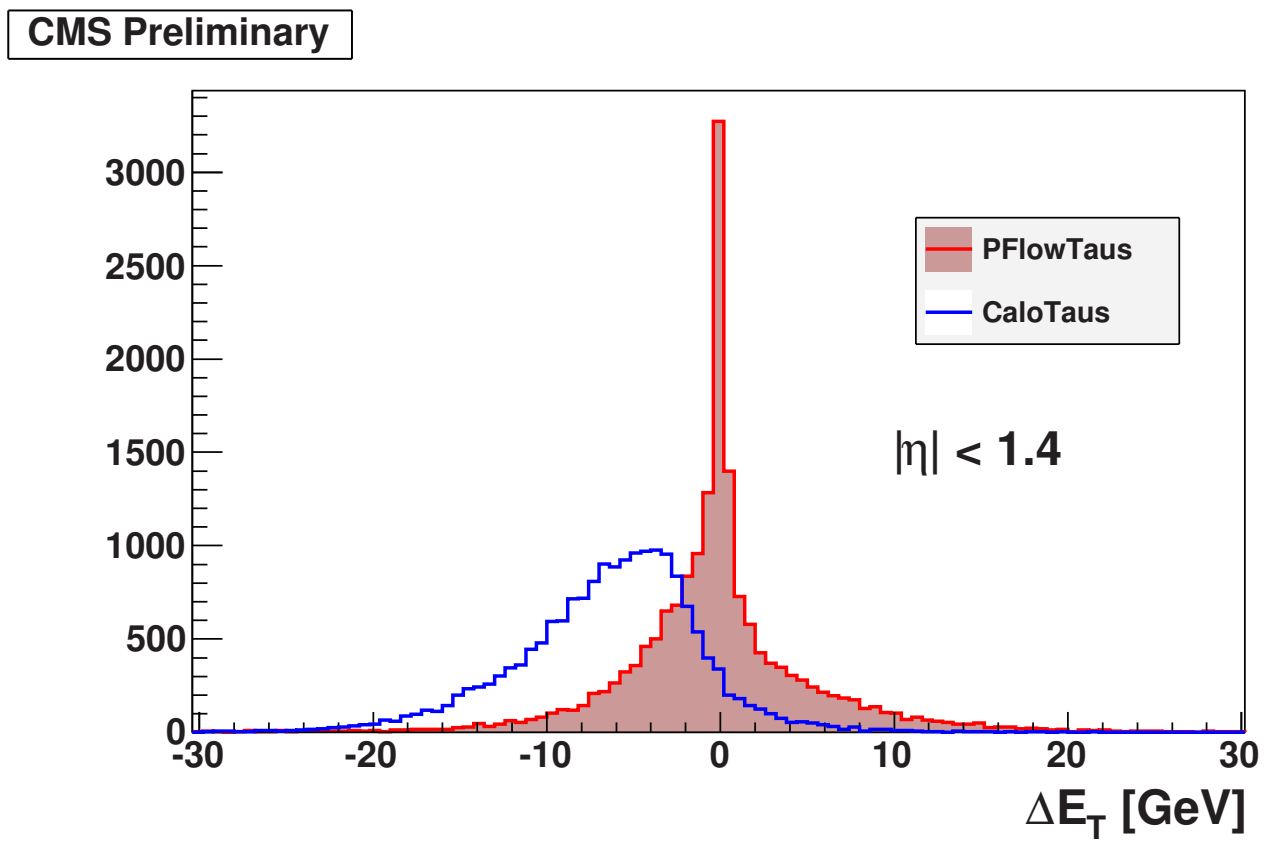

**Figure 2:** Resolution in $E_\mathrm{T}$ for hadronic $\tau$ jets from $Z$ decays, in the barrel, reconstructed using the particle flow (filled), and calorimeters (hollow).

## 2. Commissioning with the first data

A large fraction of the commissioning work was carried out with the first minimum-bias events resulting from the proton-proton collisions delivered by the LHC at a center-of-mass ($\sqrt{s}$) energy of 900 GeV[4]. It continued with the first collisions at $\sqrt{s}$=2.36 TeV. The number of high $p_\mathrm{T}$ and isolated leptons, coming from e.g. $W^\pm$ or $Z$, and high $p_\mathrm{T}$ jets, was however very limited until the LHC switched to $\sqrt{s}$=7 TeV with an increasing luminosity [5, 6].
An integrated luminosity of a few $\mathrm{nb}^{-1}$ was sufficient to check the calibration of the calorimeters. For the ECAL calibrations, photons from $\pi^0$ decays have been used. All the photons with $|\eta| < 1$ and at least 400 MeV of energy were paired. In addition the energy of the pair was required to be larger than 1.5 GeV. The resulting invariant-mass distribution of the photon pairs in the barrel can be seen in Fig. 3 where it can observed that the agreement of the average reconstructed, as obtained from a Gaussian fit from the signal is in agreement with the world average within 1%. A similar level of agreement with the Geant-based simulation is obtained.
The calibration of the hadrons is assumed to be identical for charged and neutral hadrons. Thus,

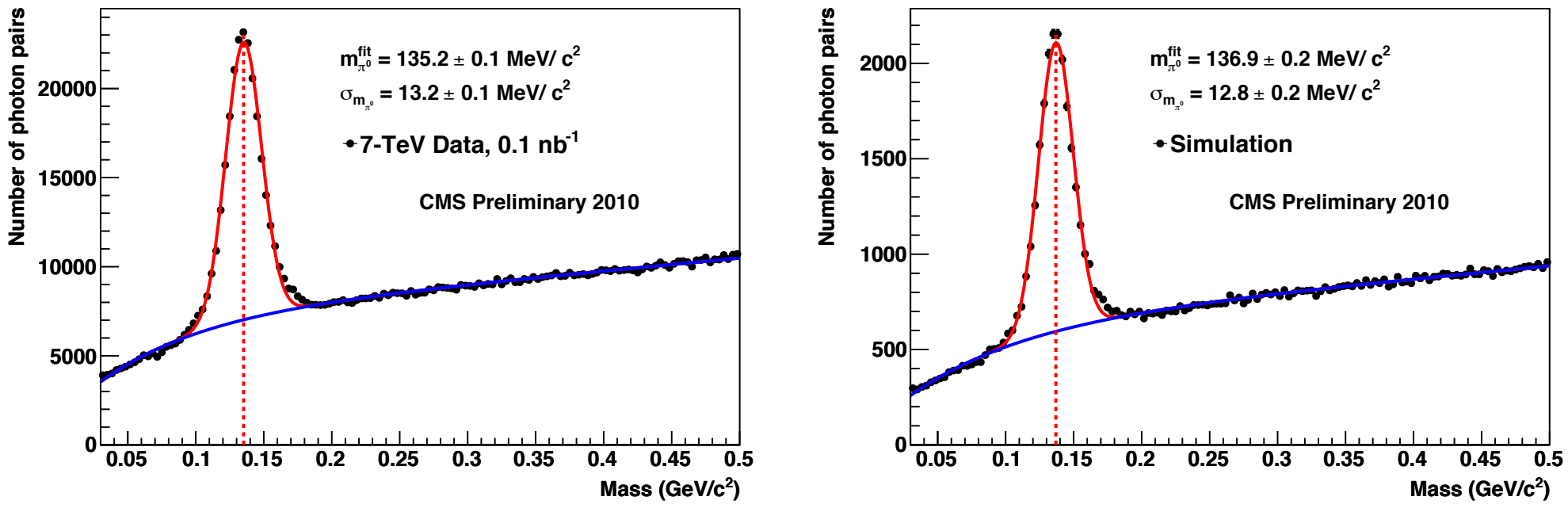

**Figure 3:** Photon-pair invariant-mass distribution in the barrel ($|\eta| < 1.0$) for the data (*left*) and simulation (*right*)

charged hadrons could be used to check the calibration. A quite pure sample of charged hadrons has been obtained by requiring high quality tracks, with $p_T > 1$ GeV/$c$, $p > 3$ GeV/$c$. In addition, the energy of the HCAL cluster had to exceed 1 GeV. Finally a loose isolation criteria was applied. The energy measured by the calorimeters could then be compared with the track momentum. The distribution thus obtained is presented in Fig. 4 and it can be observed that the slope of the calibrated energy (in red) is close to 1. For illustration, the raw energy is also presented in the same plot. The small deviation at low momentum, also present in the simulation (see [5] ) is a bias of the check, and is attributed to neutral particles overlapping with the charged hadron clusters.

Once the linking procedure and the particle calibration were proven to work in the data as expected from the simulation, it was time to compare the jet distributions. The jet multiplicity, $p_T$ and angular distributions, which can be found in [4, 5] exhibit a very satisfactory data-simulation level of agreement. The access to the jet composition, i.e. the fraction of the jet energy carried by the photons, charged and neutral hadrons in the tracker acceptance or hadronic and electromagnetic energy deposit in the forward region, was studied as a function of the jet $\eta$ (Fig. 5) and $p_T$, showing again, the goodness of the simulation. Such studies have been continuously carried out during the data taking, as illustrated in Fig. 6 extracted from [7] where the jet composition is shown as a function of the jet $p_T$ in the barrel region. The simulation reproduces well the data even with a high pile-up multiplicity.

The reconstruction was further commissioned, first by studying the $E_T^{\mathrm{miss}}$, a quantity which is known to be sensitive to detector effects. In the case of the particle-flow event reconstruction, the $E_T^{\rm miss}$ is simply the modulus of the vector sum of the reconstructed particle transverse momenta. As visible in Fig. 7, the distribution obtained in the data is reproduced by the simulation with an impressive agreement over several order of magnitudes [5]. Even more challenging, the scalar sum of the particle momenta, $\sum E_{\rm T}$, for which no noise cancellation can occur, is also faithfully reproduced by the simulation, proving the refined knowledge of the detector included in the simulation. With the large statistics available at the end of the data-taking, boosted $Z^0$ and $\gamma$+jets events can be used to study the performance of the $E_T^{\rm miss}$ reconstruction. As visible in Fig. 8, an impressive data-simulation agreement on the resolution is obtained [8].

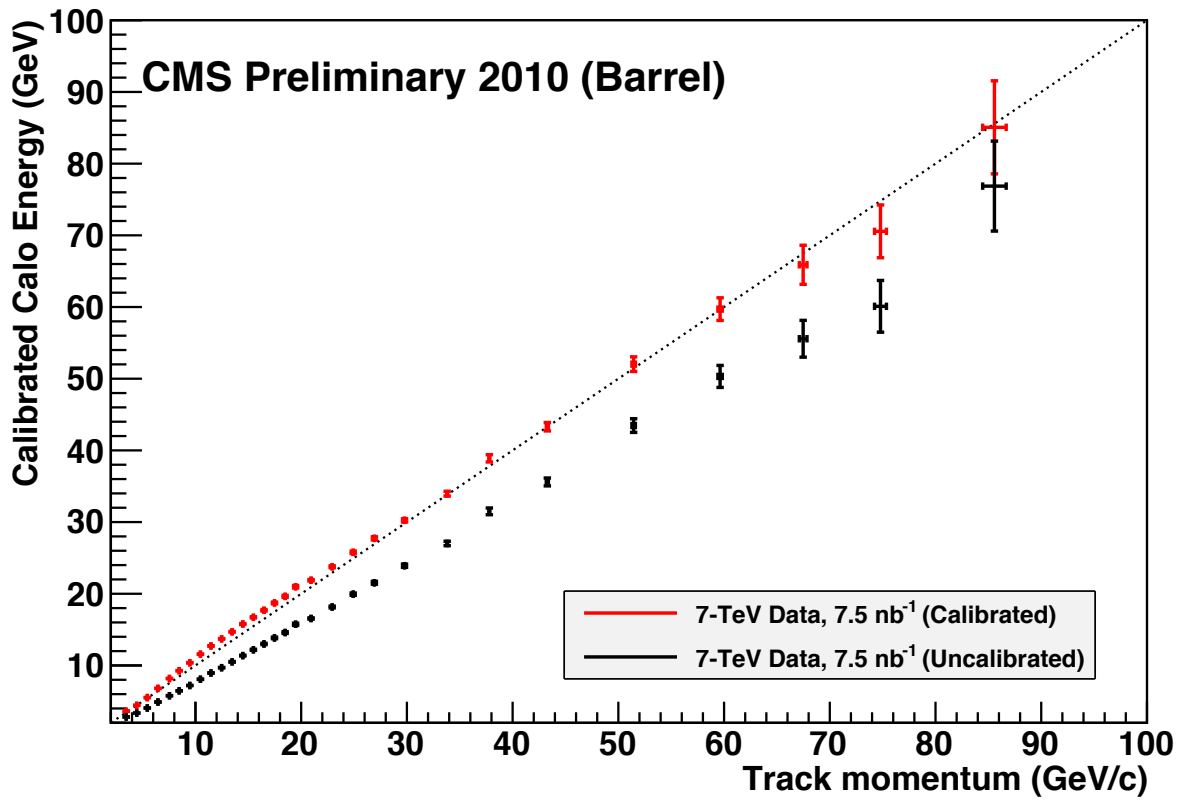

**Figure 4:** Average raw (black) and calibrated (red) calorimeter response as a function of the track momentum for charged hadrons selected in the data in the barrel.

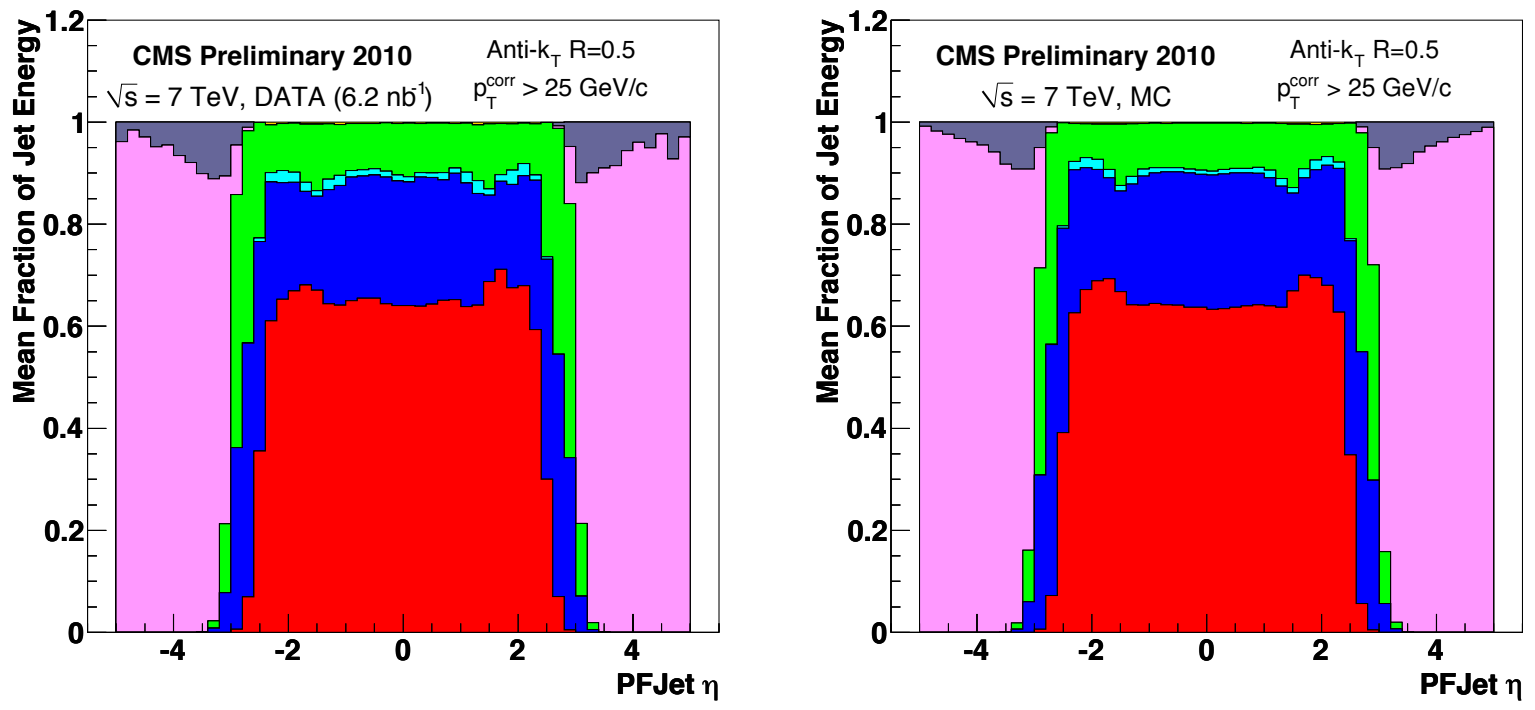

**Figure 5:** Reconstructed jet energy fractions as a function of pseudo-rapidity in the data *(left)* and in the simulation *(right)*. From bottom to top in the central region: charged hadrons, photons, electrons, and neutral hadrons. In the forward regions: hadronic deposits, electromagnetic deposits.

The reconstruction of the electrons and the muons, at high $p_{\rm T}$ and also at low $p_{\rm T}$ was, of course, also commissioned, with $J/\psi \to \ell\ell$ events (Fig. 9) as well as $W^\pm \to \ell \pm \nu$ events as described in [4, 5]. As mentioned earlier, the PF is well suited to quantify the isolation of the leptons and of the photons. The precision needed by some analyses, e.g. [9], requires clean event selections to be applied. As an example, $Z^0 \to \mu\mu\gamma$ events allowed the photon isolation around photons to be carefully monitored [10] as shown in Fig. 10 where a data-simulation agreement over more than two orders of magnitude is observed.

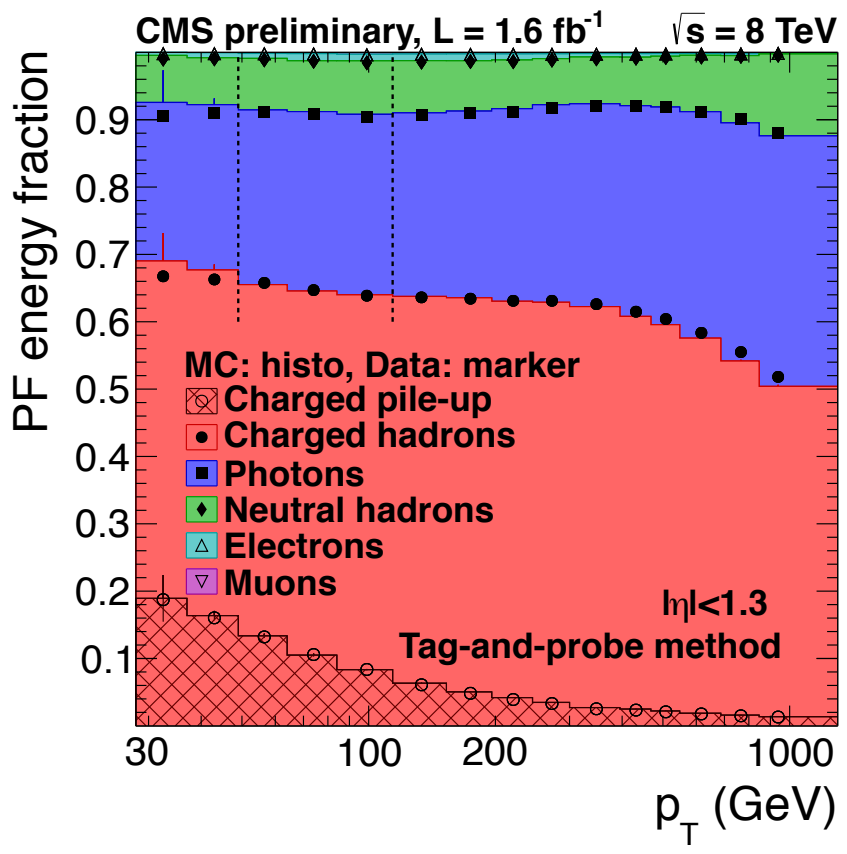

**Figure 6:** Jet composition as a function of the jet $p_{\mathrm{T}}$ for jets with $|\eta| < 1.3$ for data (histogram) and simulation (markers) .

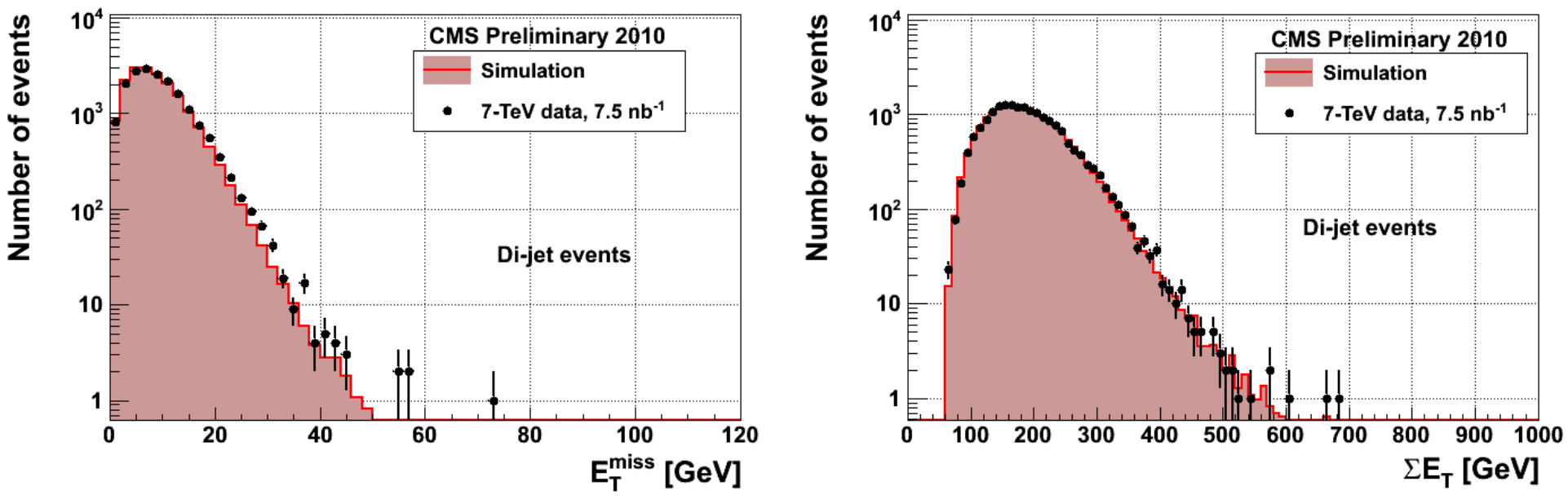

**Figure 7:** Distribution of the $E_{\mathrm{T}}^{\mathrm{miss}}$ *(left)* and $\Sigma E_T$ *(right)* for di-jet events from data (dots) and simulation (histogram).

## 3. Impact on the analyses

The gain in terms of higher efficiency, lower fake-rate, better resolution for the jet, $E_{\mathrm{T}}^{\mathrm{miss}}$ and $\tau$-lepton reconstruction, together with the reduced systematics had, obviously, a major impact on the majority of the analyses carried out in CMS since the advent of the PF. The benefits of the PF for the leptons (electrons and muons) and for the isolated photons appeared a bit later, in the period when a cutting-edge lepton and photon identification as well as the mitigation of the pile-up were crucial in the context of the Higgs boson search. Traditionally, the lepton and photon isolation are computed as the sum of the transverse momenta and energies of the tracks and deposits in the ECAL and the HCAL in a cone around the particle considered. In the PF approach, the tracks and energy deposits are simply replaced by the reconstructed particles. The gain brought by the PF isolation is illustrated in Fig. 11-left for electrons with $p_{\mathrm{T}} < 10\,\mathrm{GeV}/c$ in the endcaps. The signal and background efficiencies are relative to electrons already identified, mostly with shower-shape and track-cluster matching criteria. For the 87% efficiency working point used in the analysis, a factor of two in background rejection is achieved using the PF isolation with respect to the subdetector-based isolation. A similar background reduction is obtained for $10 < p_{\mathrm{T}} < 20\,\mathrm{GeV}/c$ electrons.

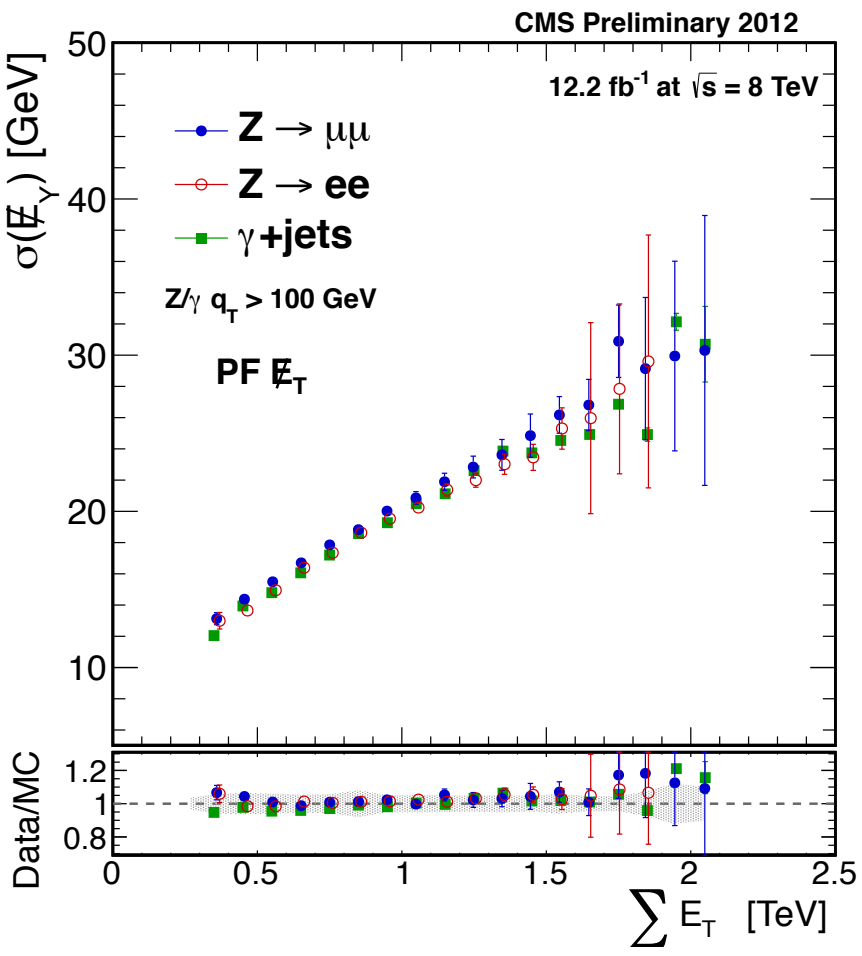

**Figure 8:** Resolution of the PF $E_{\mathrm{T}}^{\mathrm{miss}}$ projection along the y-axis as a function of $\Sigma E_T$ for events with Z and photon. Results are shown for $Z\mu\mu$ events (full blue circles), Zee events (open red circles), and photon events (full green squares).

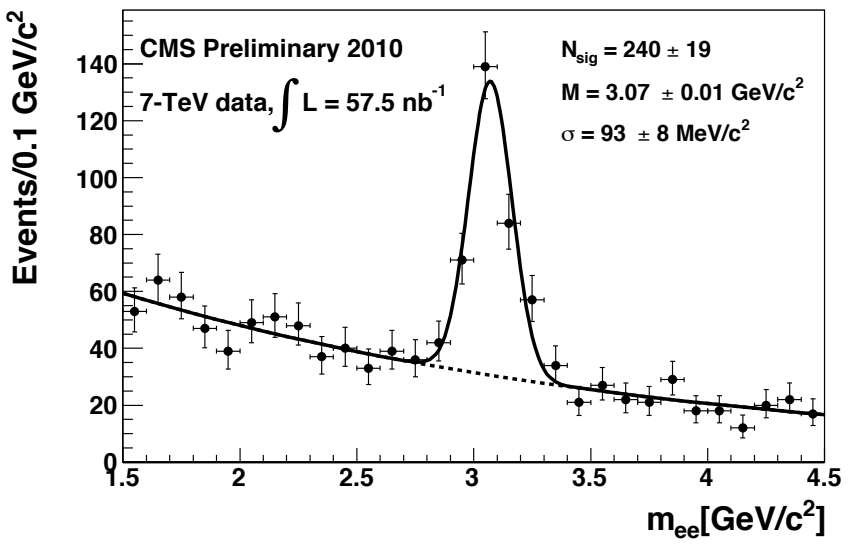

**Figure 9:** Invariant-mass distribution of the $J/\psi$ candidates decaying into an electron pairs in the first data.

The PF isolation not only gives superior performance, it also provides handles to mitigate the effect of the pile-up events. Since the charged hadrons have, by construction, a corresponding track, it is possible to impose that the charged hadrons considered for the isolation computation come from the primary vertex. As a result, the average energy contained in the isolation carried by the charged hadron exhibits little dependency on the pile-up event multiplicity, measured by the number of reconstructed vertices, as it can observed on the blue curve in Fig. 11-right. The neutrals from pile-up events falling in the isolation cannot be identified and their contribution is firstly estimated before being subtracted. Several methods exist to do so, and one of them is illustrated in Fig. 11-right where it can be seen that both the average energy density ($\rho$) and the contribution from neutrals scale linearly with the number of vertices. As a result, the average energy neutral energy caused by the pile-up events in the isolation cone is $\rho$ multiplied by a constant. The isolation of neutrals thus corrected has almost no dependency on the pile-up and so does the global PF-isolation as can be checked in Fig. 12 where the efficiency of an electron isolation cut is plotted as a function of the number of vertices. The efficiency is measured with a tag-and-probe procedure in a sample of $Z^0 \to ee$ events. The efficiency is completely stable in the barrel and decreases by about 1% in the endcaps. Once again, a spectacular data-simulation agreement is obtained.

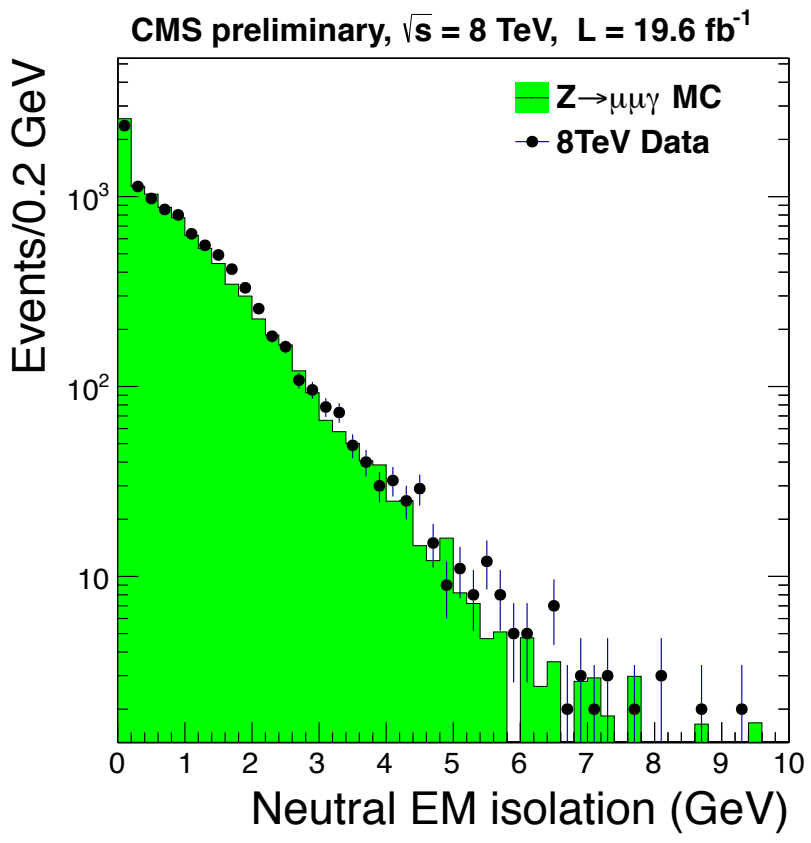

**Figure 10:** The electromagnetic isolation sum for pre-selected photons in simulated of the $Z^0 \to \mu\mu\gamma$ events compared to data for the ECAL barrel.

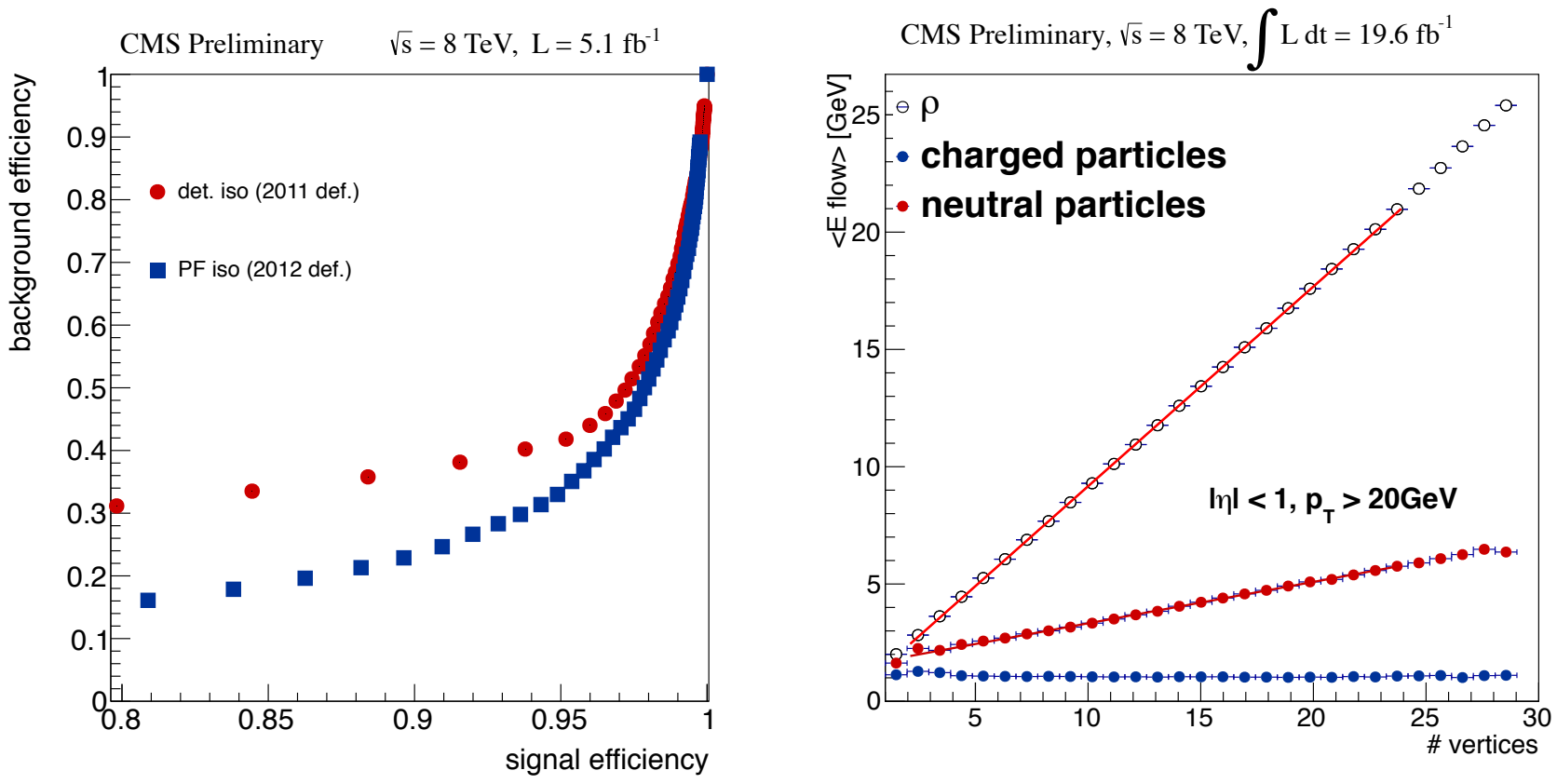

**Figure 11:** *(left)* For low $p_{\mathrm{T}}$ electrons in the endcaps, background efficiency as a function of the signal efficiency with detector isolation (red circles) and PF isolation (blue squares).*(right)* For electrons in the central region with $p_{\mathrm{T}} > 20\,\mathrm{GeV}/c$: evolution of the average event energy density (open circles) and energy within the isolation cone carried by the charged particles (blue circles) and neutral particles (red circles).

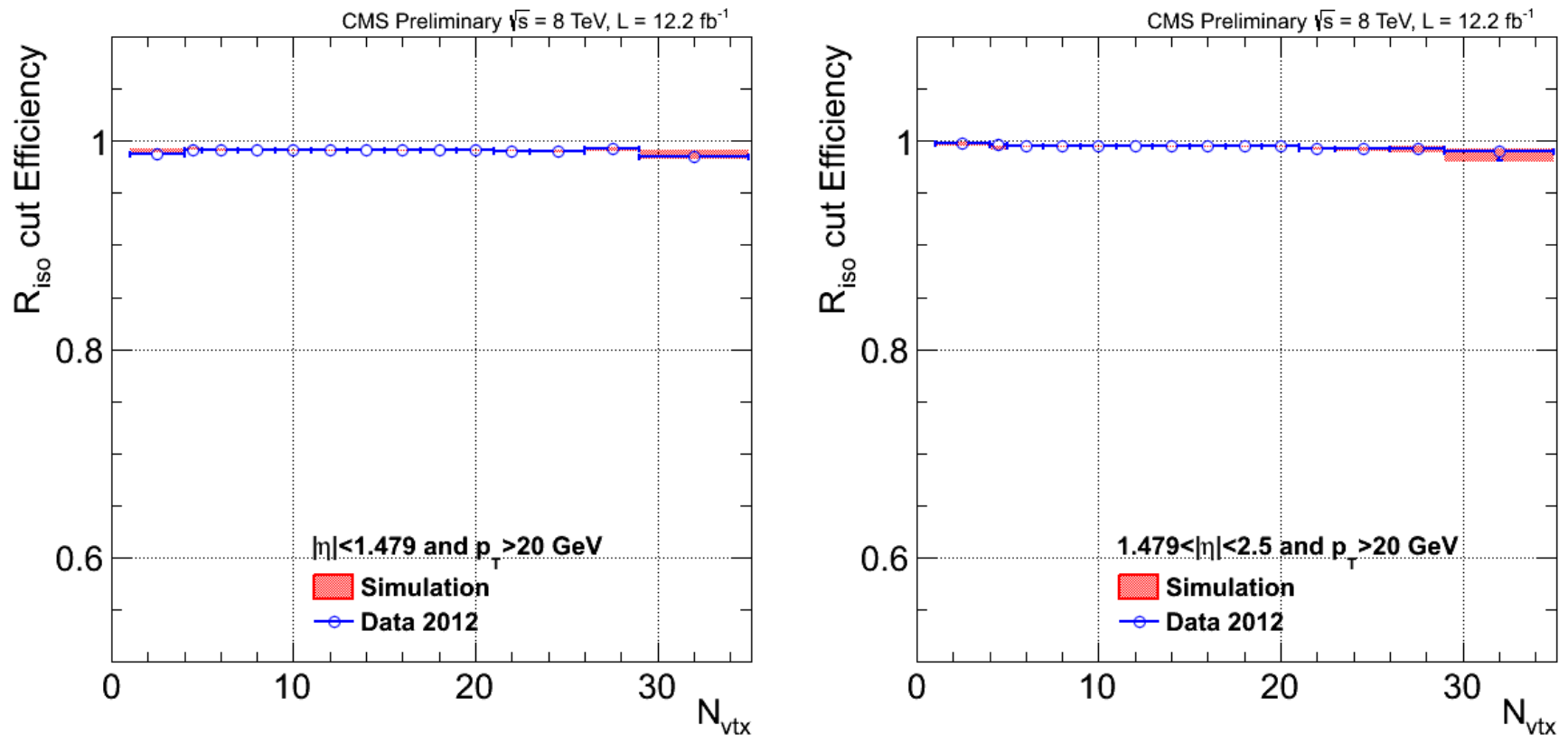

**Figure 12:** Efficiency of the isolation requirement for identified electrons as obtained with a tag-and-probe method with $Z^0 \to$ ee events in the barrel *(left)* and in the endcaps *(right)*.

## 4. Conclusion

The CMS detector is well suited for an efficient PF algorithm. The PF has been deployed in CMS and is currently used in most of the analyses. Even though the concept is simple, the implementation required some work. The performance of jets, $E_{\mathrm{T}}^{\mathrm{miss}}$ reconstruction, lepton reconstruction and identification have been significantly improved. Most of the key ingredients of the algorithm have been commissioned with the very first data, in a short timescale, demonstrating the accuracy of the CMS simulation from the beginning. The Particle Flow has not only proven to work well with high pile-up conditions, it appears as the only way to preserve and even improve the performance of CMS for the next rounds of data-taking.

## Acknowledgments

I would like to thank the organizers for the very pleasant and interesting conference.